\documentclass[prl,twocolumn,showpacs,amsmath,amssymb,superscriptaddress]{revtex4}

\usepackage{graphicx}
\usepackage{dcolumn}
\usepackage{bm}

\begin{document}

\title{
Superfluid state of repulsively interacting three-component fermionic atoms in optical lattices
}

\author{Kensuke Inaba}
\affiliation{NTT Basic Research Laboratories, NTT Corporation, Atsugi 243-0198, Japan}
\affiliation{JST, CREST, Chiyoda-ku, Tokyo 102-0075, Japan}
\author{Sei-ichiro Suga}
\affiliation{Department of Materials Science and Chemistry, University of Hyogo, Himeji 671-2280, Japan}

\date{\today}

\begin{abstract}
We investigate the superfluid state of repulsively interacting three-component (color) fermionic atoms in optical lattices.
When the anisotropy of the three repulsive interactions is strong, atoms of two of the three colors form Cooper pairs and atoms of the third color remain a Fermi liquid. An effective attractive interaction is induced by density fluctuations of the third-color atoms. 
This superfluid state is stable against changes in filling close to half filling. We determine the phase diagrams in terms of temperature, filling, and the anisotropy of the repulsive interactions. 
\end{abstract}


\preprint{APS/123-QED}

\maketitle

The cold atoms in optical lattices offer a new way of studying fascinating aspects of quantum many-body physics \cite{Greiner2008,Bloch2005,Jaksch2005}.
High controllability of the atoms allows us to create novel systems, which extends our current knowledge in, {\it e.g.}, condensed matter physics. 
The research on Bose-Fermi mixtures trapped in optical lattices extends the paradigm of the Mott transition \cite{Sugawa2011}. 
Fermionic atoms with odd numbers of internal degrees of freedom are other examples. 
Recently, interesting features of degenerate three-component fermionic gases were actively investigated \cite{Ottenstein2008,Huckans2009,Miyatake2010}. 
For three-component (color) fermionic atoms in optical lattices, it was shown theoretically that the paired Mott insulator (PMI) can appear in spite of half filling, 
where the average atom number per site is the non-integer 3/2 \cite{Inaba2010b,suga2011}. 
When two of the three repulsions between atoms with different colors are stronger than the other,  pairs of weakly repulsing atoms are formed to avoid the two stronger repulsions. 
Therefore, in the PMI the effective particle number at a site becomes an integer \cite{Inaba2010b}. 
This feature implies that, in the Fermi liquid (FL) states in the vicinity of the PMI transition point at half filling and close to half filling, superfluid pair fluctuations are strongly enhanced at low temperatures, although the bare interactions are repulsive. 
We expect a characteristic superfluid (SF) state to appear there. In this SF, atoms of two of the three colors form Cooper pairs and atoms of the third-color remain a FL. 
This SF state is similar to the color superfluid of the attractively interacting three-component systems \cite{Rapp2007,Inaba2009a,Inaba2011}. 
Knowledge of this novel SF provides an insight into the Cooper pairing mechanism of other repulsively interacting systems in, for instance,  condensed matter physics.

In this Letter, we investigate the SF transition in  repulsively interacting three-component fermionic atoms in optical lattices. 
We first discuss the effective interactions using Feynman diagrammatic approaches. 
By combining the dynamical mean-field theory (DMFT) with a perturbative approach \cite{Georges1996}, we then investigate the SF properties in detail. 
We show that the SF phase appears close to the PMI phase, when the anisotropy of the interactions is strong. 
We reveal that the effective attractive interaction that forms Cooper pairs of two-color atoms is mediated by density fluctuations of the third-color atoms. 
We also perform unperturbative calculations  using the self-energy functional approach \cite{Potthoff2003}. 
The two sets of results are complementary.

The low-energy properties of the present system are well described by the following Hubbard Hamiltonian: 
\begin{eqnarray}
\hat{\cal H}=-t \sum_{\langle i,j \rangle}\sum_{\alpha=1}^{3}
       \hat{a}^\dag_{i\alpha} \hat{a}_{j\alpha} 
  &-& \sum_{i}\sum_{\alpha=1}^{3} \mu_\alpha \hat{n}_{i \alpha} \nonumber \\
  &+& \frac{1}{2}\sum_{i}\sum_{\alpha\not=\beta} 
       U_{\alpha\beta} \hat{n}_{i \alpha} \hat{n}_{i \beta},   
\label{eq_model}
\end{eqnarray}
where the subscript $\langle i,j \rangle$ is the summation over the nearest-neighbor sites, and $\hat{a}^\dag_{i\alpha} (\hat{a}_{i\alpha})$ and $\hat{n}_{i\alpha}$ are creation (annihilation) and number operators of a fermion with color $\alpha$ at the $i$th site. 
Here, $t$ denotes the hopping integral. 
We choose the chemical potential $\mu_\alpha$  to satisfy the balanced population of each color atoms; $n(=\langle n_\alpha\rangle)$.
For simplicity, we set $U_{12}\equiv U$ and $U_{23}=U_{31}\equiv U'$, and $\mu_1=\mu_2\equiv \mu$ and $\mu_3\equiv\mu'$. 
We introduce an anisotropy ratio $R\equiv U/U'$. 
We neglect the confinement potential for the first approximation. 
We also neglect the symmetry breaking phases except for the superfluid state.

\begin{figure}[t]
\includegraphics[width=58mm]{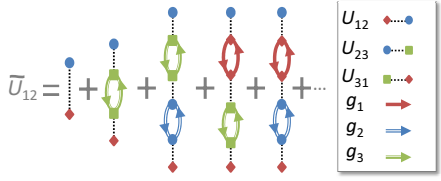}
\caption{(Color Online) Diagrammatic representation of the effective interaction between color-1 and 2 atoms. 
}
\label{fig_rpa}
\end{figure}

Let us first consider the effective interaction ($\tilde{U}_{12}$) between color-1 and 2 atoms by adopting a random phase approximation analysis as shown in Fig. \ref{fig_rpa}. 
The contribution of all the bubble diagrams is summarized in the following expression,  
\begin{eqnarray}
\tilde{U}_{12}({\bf x})&=&\Big[U_{12}-U_{13}\chi_3({\bf x})U_{23}\Big]
\bigg[1 -U_{12}^2\chi_1({\bf x})\chi_2({\bf x}) \nonumber \\
&&-U_{13}^2\chi_1({\bf x})\chi_3({\bf x})  -U_{23}^2\chi_2({\bf x})\chi_3({\bf x}) \nonumber \\
&&+2U_{12}U_{23}U_{13}\chi_1({\bf x})\chi_2({\bf x})\chi_3({\bf x})\bigg]^{-1},
\end{eqnarray}
where ${\bf x}=(i\nu_l,{\bm k})$, $\nu_l$ is the bosonic Matsubara frequency, and ${\bm k}$ is the wave vector. 
The bubble diagram corresponding to density fluctuations is given by $\chi_\alpha({\bf x})=-\sum_{\bf y}g_\alpha({\bf y})g_\alpha({\bf y+x})$, 
where ${\bf y}=(i\omega_n,{\bm q})$ with $\omega_n$ and ${\bm q}$ being the fermionic Matsubara frequency and the wave vector, respectively, and the bare Green's function of color-$\alpha$ atoms is denoted by $g_\alpha({\bf y})=1/(i\omega_n+\mu_\alpha-\varepsilon_{\bm q})$. 
The other effective interactions are evaluated in the same way. 
Since $U_{12}\equiv U$ and $U_{13}=U_{23}\equiv U'$, we set $\chi_1=\chi_2 \equiv \chi$ and $\chi_3 \equiv \chi'$. 
The effective interactions $\tilde{U}({\bf x})(\equiv\tilde{U}_{12}({\bf x}))$ and $\tilde{U}'({\bf x})(\equiv\tilde{U}_{23}({\bf x})=\tilde{U}_{31}({\bf x}))$ are given by 
\begin{eqnarray}
\tilde{U}({\bf x})&=&\frac{U-U'^2\chi'({\bf x})}
{[1-U\chi({\bf x})][1+U\chi({\bf x})-2U'^2\chi({\bf x})\chi'({\bf x})]}, 
\label{U}\\
\tilde{U}'({\bf x})&=&\frac{U'}
{1+U\chi({\bf x})-2U'^2\chi({\bf x})\chi'({\bf x})}. 
\label{U'}
\end{eqnarray}
For $U \ll U' < t$ ($R\ll1$), $\tilde{U}({\bf x})\sim U -U'^2\chi'({\bf x})$. The second term can be regarded as the effective attraction. When the effective attraction overcomes the bare interaction $U$, $\tilde{U}({\bf x})$ becomes attractive. 
By contrast, for $U=U'< t$ ($R=1$) we obtain the effective repulsive interactions $\tilde{U}({\bf x})=\tilde{U}'({\bf x}) \sim U$. 
This simple discussion suggests that when the anisotropy is strong ($R \ll 1$) the effective attractive interaction between color-1 and 2 atoms is induced by density fluctuations of the color-3 atoms. 
Furthermore, for $R \ll 1$ the FL-PMI transition occurs in the strongly correlated region as shown in Ref. \cite{Inaba2010b}. 
Therefore, we can expect  the $s$-wave SF caused by the attractive $\tilde{U}({\bf x})$ to appear in the FL phase close to the PMI transition point at low temperatures.
We perform DMFT calculations to confirm this expectation.
When the system approaches the PMI transition point, local correlation effects enhance the quantum fluctuations. In particular, SF pair fluctuations between color-1 and 2 atoms are expected to be strongly enhanced. To deal precisely with the local correlation effects we apply the infinite-dimensional system, which is a nontrivial limit for discussing both the Mott transition and the s-wave superfluidity \cite{Georges1996,Garg2005}.
For noninteracting atoms, we use a semicircular density of states, $\rho_0(x)=\sqrt{4t^2-x^2}/(2\pi t^2)$. 
\begin{figure}[t]
\includegraphics[width=65mm]{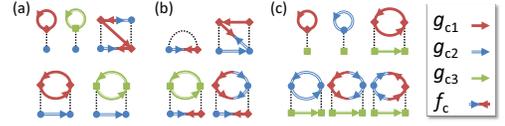}
\caption{(Color Online) 
Diagrams for the (a) normal and (b) anomalous self-energies of color-1 and 2 atoms, respectively, and (c) the normal self-energies of color-3 atoms. }
\label{fig_diag}
\end{figure}
To investigate the SF, we carry out the second-order perturbative expansion of the self-energy within the DMFT framework. 
This theory is namely the iterated perturbation theory (IPT)  \cite{Georges1996}. 
By using the Nambu spinor representation $\hat{\psi}_{\bm q}=(c_{{\bm q}1}, c_{-{\bm q}2}^\dag, c_{{\bm q}3})$, the Green's function matrix is given by
\begin{eqnarray}
\boldsymbol{G}({\bf y}) = 
\left(\begin{array}{ccc}
G_1({\bf y}) &F({\bf y})&0\\
F({\bf y})&-G_1(-{\bf y})&0\\
0&0&G_3({\bf y})
\end{array}\right).
\end{eqnarray}
In the infinite-dimensional system, the self-energy matrix is purely local and independent of ${\bm q}$, yielding $\bm{\Sigma}({\bf y})=\bm{\Sigma}(i\omega_n)$, where 
\begin{eqnarray}
\boldsymbol{\Sigma}(i\omega_n) = 
\left(\begin{array}{ccc}
\Sigma_1(i\omega_n) &S(i\omega_n)&0\\
S(i\omega_n)&-\Sigma_1(-i\omega_n)&0\\
0&0&\Sigma_3(i\omega_n)
\end{array}\right).
\end{eqnarray}
According to the DMFT \cite{Georges1996}, $\boldsymbol{\Sigma}(i\omega_n)$ can be obtained from the effective impurity model, and 
the impurity Green's function matrix ${\bm G}_{\rm imp}(i\omega_n)$ must be self-consistently equal to the local Green's function matrix of the original system, ${\bm G}_{\rm loc}(i\omega_n)[=\sum_{\bm q}{\bm G}({\bf y})]$.

We calculate the first- and second-order self-energies, $\boldsymbol{\Sigma}^{(1)}$ and $\boldsymbol{\Sigma}^{(2)}$, respectively, by collecting the diagrams shown in Fig. \ref{fig_diag}. 
Here, the arrowed lines denote the diagonal ($g_{c\alpha}$) and off-diagonal ($f_c$) components of the cavity Green's function matrix ${\bm g}_c$, which is the noninteracting impurity Green's function matrix and satisfies the Dyson equation, ${\bm g}_c=[{\bm G}_{\rm imp}^{-1}+\boldsymbol{\Sigma}]^{-1}$. 
To obtain  precise results close to the PMI transition point, we modify the self-energy matrix as $\boldsymbol{\Sigma}=\boldsymbol{\Sigma}^{(1)}+{\bm A}\boldsymbol{\Sigma}^{(2)}$ according to Ref. \cite{Garg2005}. 
The matrix ${\bm A}$ is determined so as to reproduce the exact $1/\omega$ expansion of ${\bm G}_{\rm imp}$.
The diagonal elements are  $A_{11}=A_{22}=[R^2n_2(1-n_2)+n_3(1-n_3)-R^2|\Phi|^2+2R(D_{23}-n_2n_3)]/[R^2 n_{c2}(1-n_{c2})+n_{c3}(1-n_{c3})-R^2|\Phi_c|^2]$, 
$A_{33}=[n_1(1-n_1)+D_{12}-n_1n_2+|\Phi|^2]/[n_{c1}(1-n_{c1})+|\Phi_c|^2]$, and the off-diagonal elements are zero. 
The number of atoms $n_\alpha$ and the SF order parameter $\Phi\equiv\langle \hat{a}_{i1}\hat{a}_{i2}\rangle$ are calculated from the diagonal and off-diagonal elements of 
$\sum_n{\bm G}_{\rm imp}(i\omega_n)e^{i\omega_n0^+}$, respectively.
In the same way, $n_{c\alpha}$ and $\Phi_c$ are calculated from the cavity Green's function. 
The double occupancy $D_{\alpha\beta}=\langle n_\alpha n_\beta \rangle$ is obtained from $D_{\alpha\beta}=-{\rm Tr} \left[{\bm G}_{\rm imp}\partial\boldsymbol{{\Sigma}}/\partial U_{\alpha \beta} \right]$.

\begin{figure}[t]
\includegraphics[width=8cm]{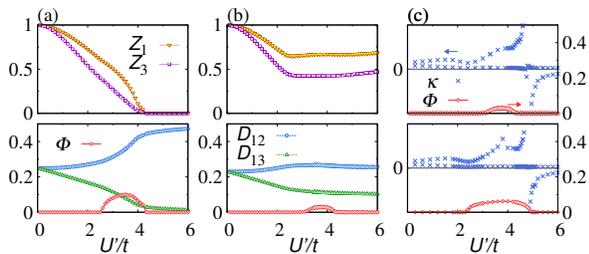}
\caption{(Color Online) 
(a) and (b) Renormalization factor $Z_\alpha$, SF order parameter $\Phi$, and double occupancy $D_{\alpha\beta}$ calculated with the IPT as functions of $U'/t$ at $R=0.1$ and $T/t=0.03$ for fillings (a) $n=0.5$ and (b) $n=0.48$. 
(c) Eigenvalues ($\kappa$) of $\kappa_{ab}$ at $T/t=0.03$ (top) and 0.02 (bottom) for $R=0.1$ and $n=0.48$. 
Negative values denote the instability of the homogeneous phases. 
For comparison, $\Phi$ is also plotted. 
}
\label{fig_udep}
\end{figure}
\begin{figure}[t]
\includegraphics[width=7cm]{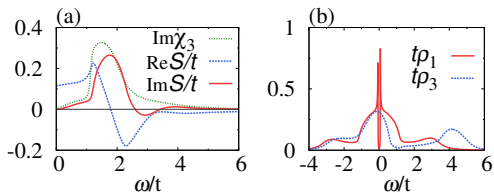}
\caption{(Color Online) 
(a) Anomalous self-energy ${\rm Re}S(\omega_+)$, ${\rm Im}S(\omega_+)$, and imaginary part of the polarization function ${\rm Im}\chi_3(\omega_+)$. (b) Single-particle excitation spectra $\rho_\alpha(\omega)$. These quantities are calculated with the IPT.  Parameters are $U'/t=3$, $R=0.1$, $n=0.48$, and $T/t=0.01$.
}
\label{fig_dos}
\end{figure}
By using the IPT, we calculate $\Phi$, $D_{\alpha \beta}$, and the renormalization factor $Z_\alpha\equiv 1/(1-\partial \Sigma_\alpha/\partial \omega)|_{\omega=0}$.
Figure \ref{fig_udep}(a) and (b) show the $U'/t$ dependence of these quantities at  $R=0.1$ and $T/t=0.03$ for different fillings $n=0.5$ (half filling) and $n=0.48$, respectively. 
In Fig. \ref{fig_udep}(a), $Z_\alpha$ decreases with increasing in $U'/t$ and vanishes at $U'/t\sim 4.2$. 
The double occupancy $D_{12}$ ($D_{13}$) increases (decreases) as $U'/t$ increases. These results indicate that the PMI transition occurs to avoid the stronger $U'/t$. 
When $U'/t$ approaches the PMI transition point from below, the SF order parameter becomes finite and reaches zero with a tiny jump at the transition point, suggesting a discontinuous transition. 
In contrast, $\Phi$ increases smoothly at $U'/t\sim2.5$ with increasing $U'/t$.
The transition between the FL and SF is thus continuous. 
For $n=0.48$ the PMI does not appear as shown in Fig. \ref{fig_udep}(b), because the effective particle number at a site is non-integer \cite{Inaba2010b,Gorelik2009}. 
Note that the density-wave states are also unstable, unless at least two of three color atoms are at commensurate filling \cite{Miyatake2010,HH2004,Rosch}. 
On the other hand, the SF phase appears in $3<U'/t<4.2$ for $n=0.48$. 
We thus expect this characteristic SF to be stable near half-filling for the balanced population.

We further examine the stability of the above SF phase against the phase separation (PS), which was pointed out in recent theoretical studies on the attractively interacting three-component systems \cite{Titv,Priv}. 
We consider the possibility of the PS between paired color-1 and 2 atoms, and unpaired color-3 atoms. To this end, we calculate the $2\times2$ compressibility matrix $\kappa_{ab}\equiv \partial N_a/\partial\mu_b$ with $N_a=n_1+n_2, n_3$ and $\mu_b=\mu,\mu'$. 
Figure \ref{fig_udep}(c) exhibits the eigenvalues of $\kappa_{ab}$ at $T/t=0.02$ and 0.03 for $R=0.1$ and $n=0.48$. 
We find that for both $T/t=0.02$ and $0.03$ one of the eigenvalues shows anomaly at $U'/t \sim 4.8$ and becomes negative in $U'/t \gtrsim 4.8$, which indicates the instability to the PS. In contrast, the stable homogeneous SF and FL phases are found in $U'/t \lesssim 4.8$. 

We now investigate the SF in terms of its dynamical properties. 
As discussed concerning Eq. (\ref{U}), density fluctuations of color-3 atoms play a key role in the appearance of the SF. 
Therefore, we discuss ${\rm Im}\chi_3(\omega)$ and the dynamics of the anomalous self-energy $S(\omega)$. 
In Fig. \ref{fig_dos}(a), we show ${\rm Im}\chi_3(\omega_+)$, ${\rm Re}S(\omega_+)$, and ${\rm Im}S(\omega_+)$ with $\omega_+\equiv\omega+i0^+$ for $U'/t=3$, $R=0.1$, $n=0.48$, and $T/t=0.01$. 
The spectra ${\rm Im}\chi_3(\omega)$ have a peak around $\omega_p/t\sim1.6$, indicating that the scattering caused by the density fluctuations of color-3 atoms occurs prominently around $\omega\sim\omega_p$. 
The spectra ${\rm Re}S(\omega_+)$ exhibit resonant behavior and change their sign at $\omega/t\sim1.8$, which is larger than $\omega_p/t$. 
At $\omega/t\sim1.8$, ${\rm Im}S(\omega_+)$ has a peak. 
Since ${\rm Re}S(\omega_+)$ is a measure of the effective interaction \cite{Scalapino}, the effective interaction between color-1 and 2 atoms is attractive when $\omega/t<1.8$.
The frequency of the peak of ${\rm Re}S(\omega_+)$ is smaller than $\omega_p/t$, and ${\rm Re}S(\omega_+)$ remains negative for a large $\omega/t$. 
These features are caused by the bare repulsion $U$ between color-1 and 2 atoms \cite{Scalapino}. 
These dynamical properties are qualitatively the same as those of the strong-coupling superconductor caused by the electron-phonon interaction \cite{Scalapino}.

In Fig. \ref{fig_dos}(b), we show  single-particle excitation spectra defined as $\rho_\alpha(\omega)=-(1/\pi){\rm Im}G_{\rm imp, \alpha}(\omega_+)$.
The $s$-wave superfluid gap is seen around the Fermi energy ($\omega=0$) in $\rho_1(\omega)[=\rho_2(\omega)]$, while the renormalized peak is seen in $\rho_3(\omega)$ owing to the FL nature. 
This $s$-wave spectral gap is consistent with the dynamical properties shown in Fig. \ref{fig_dos}(a), which are qualitatively the same as those of the strong-coupling phonon-mediated superconductor.
We find that the incoherent Hubbard band structures appear in both spectra $\rho_1(\omega)$ and $\rho_3(\omega)$, which stem from strong correlation effects. 
Actually, the peak position of ${\rm Im}S(\omega_+)$ is in this incoherent Hubbard band region of $\rho_1(\omega)$. 
On the basis of the results described so far, we can argue that the effective attractive interaction between color-1 and 2 atoms is mediated by the density fluctuations of color-3 atoms.

\begin{figure}[t]
\includegraphics[width=8.5cm]{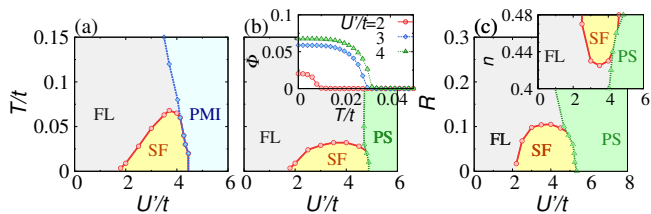}
\caption{(Color Online) Phase diagrams obtained with the IPT. 
$U'-T$ phase diagrams for $R=0.1$ and different fillings (a) $n=0.5$ and (b) $0.48$. Inset in (b) shows $\Phi$ as a function of $T/t$ at $R=0.1$ for $n=0.48$. 
(c) Phase diagrams for $T/t=0.03$. 
Main panel and inset are $U'$-$R$ and $U'$-$n$ phase diagrams for $n=0.48$ and $R=0.1$, respectively. 
}
\label{fig_phase}
\end{figure}
By systematically changing $T$ and $U'$, we determine the finite-temperature phase diagrams for $n=0.5$ and $0.48$ at $R=0.1$. 
The comparison of these two phase diagrams is very informative, although we neglect the density-wave state at $n=0.5$. 
For $n=0.5$, the SF phase appears adjacent to the PMI phase as shown in Fig. \ref{fig_phase}(a). 
For $n=0.48$, the SF phase extends to a stronger $U'/t$ region than that for $n=0.5$ as shown in Fig. \ref{fig_phase}(b), suggesting that SF pair fluctuations survive in the vicinity of the PMI phase. 
However, the homogeneous phase becomes unstable in $U'/t \gtrsim 4.8$. 
The inset in Fig. \ref{fig_phase}(b) shows the temperature dependence of $\Phi$ for $n=0.48$ and $R=0.1$. 
We find $\Phi \propto \sqrt{T-T_c}$ with $T_c$ being the SF transition temperature, which is typical of  weak- and strong-coupling phonon-mediated superconductors \cite{Scalapino}.
We also determine the $R$- and $n$-dependent phase diagrams shown in Fig. \ref{fig_phase}(c). 
From Eq. (\ref{U}) we expect $\Phi$ to decrease and then vanish with increasing $R$, because the effective attractive interaction cannot overcome the bare repulsion $U$ in the weak anisotropy (large $R$) region. 
We find that the SF appears in $R \lesssim0.11$  as shown in the main panel of Fig. \ref{fig_phase}(c), suggesting that the strong anisotropy is thus necessary for the appearance of the SF. 
We remark that as $n$ deviates from 0.5, $\Phi$ decreases monotonically due to the suppression of density fluctuations of color-3 atoms. 
We find that the SF survives down to $n\sim0.43$ as shown in the inset of Fig. \ref{fig_phase}(c).

Next, we discuss the SF transition  using an unperturbative numerical method.
The higher-order contributions, which are not included in our diagrammatic analyses, may play an important role. 
We thus perform the self-energy functional approach calculations \cite{Potthoff2003}. 
This method was successfully used for studying the attractively interacting three-component Hubbard model \cite{Inaba2009a,Inaba2011}. 
In this method we introduce a proper reference system, which has to include the same interacting Hamiltonian as that of the original Hubbard Hamiltonian. 
We employ the two-site Anderson impurity model as the reference system. 
The local correlation effects are unperturbatively taken into account via the reference self-energy calculated by the exact diagonalization method \cite{Potthoff2003,Inaba2009a,Inaba2011}. 
\begin{figure}[t]
\includegraphics[width=8.8cm]{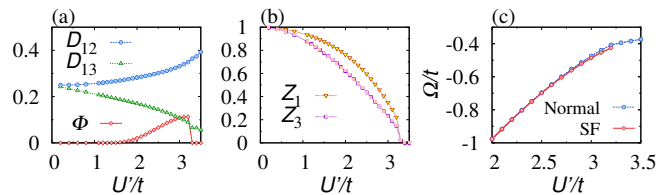}
\caption{(Color Online) 
(a) SF order parameter $\Phi$, Double occupancy $D_{\alpha\beta}$, 
(b) renormalization factor $Z_\alpha$, and
(c) grand potentials of the SF and the normal states (FL and PMI). 
These quantities are calculated using the self-energy functional approach for $R=0.1$, $T=0$, and $n=0.5$. 
}
\label{fig_sfa}
\end{figure}
In Fig. \ref{fig_sfa}(a) and (b), we show $\Phi$, $D_{\alpha\beta}$, and $Z_\alpha$ for $n=0.5$, $R=0.1$, and $T=0$. 
The $U'/t$ dependences of these quantities are qualitatively the same as those obtained by the IPT. 
In Fig. \ref{fig_sfa}(c), we compare the grand potentials $\Omega$ of the SF and the normal state obtained by the calculations without any symmetry breaking. 
Note that the normal state in $U'/t<3.2$ is the FL, while that in $U'/t>3.2$ is the PMI. 
The grand potential of the SF is lower than that of the FL. 
We thus confirm that the SF is stable. 
The qualitative agreement between the IPT results and the present unperturbative calculations means that our IPT treatment captures the essentials of the correlation effects.

Finally, we discuss the relation between our theoretical results and the experiments. 
The double occupancy shown in Fig. \ref{fig_udep}(a) and (b), and the single-particle excitation spectra shown in Fig. \ref{fig_dos}(b) can be measured by state-of-the-art experimental techniques \cite{Sugawa2011,JILA}. Atoms with convenient magnetic Feschbach resonances such as $\rm ^6Li$ are good candidates for realizing the SF, because the anisotropy of the interactions is controllable. 
It was pointed out that, for attractively interacting $\rm ^6Li$ atoms in optical lattices, the strong three-body loss makes the homogeneous phases unstable \cite{Titv}. However, as shown in the experiment \cite{Ottenstein2008}, the loss rate in the repulsive region is significantly small as compared with that in the attractive region. 
Furthermore, the three-body loss effects are suppressed in the repulsively interacting system, because the two-body repulsions decrease the triple occupancy $\langle n_1n_2n_3\rangle$. 
Accordingly, the loss-induced PS can be negligible and the phase diagrams shown in Fig. \ref{fig_phase} are adequate for the repulsively interacting three-component $\rm ^6Li$ atoms in optical lattices.

In summary, we have pointed out that the SF appears in repulsively interacting three-component fermionic atoms in optical lattices. 
We have shown that the Cooper-pairing mechanism is basically similar to that of the conventional phonon-mediated superconductivity, while the SF phase appears close to the PMI phase.
When we consider the realistic situations, we have to take account of the confinement potential and the effects of the small but finite loss rate depending on species of atoms. These issues are subjects for a future study.

\begin{acknowledgments}
We thank T. Ohashi for useful comments and valuable discussions. 
This work was supported by JSPS KAKENHI (23540467) and MEXT KAKENHI (21104514). 
\end{acknowledgments}

\end{document}